\documentclass{article}
\usepackage{spconf,amsmath,graphicx}
\usepackage{amsfonts}
\usepackage{gensymb}
\usepackage{algorithmic}
\usepackage{graphicx}
\usepackage{textcomp}
\usepackage{xcolor}
\usepackage{hyperref}
\usepackage{subfigure}
\usepackage[bottom]{footmisc}

\title{Improving dual-microphone speech enhancement by learning cross-channel features with multi-head attention}
\name{Xinmeng Xu$^{1}$\quad Rongzhi Gu$^{1}$ \quad Yuexian Zou$^{1,2,*}$ \thanks{$^*$Corresponding author}\thanks{The research work is supported by Shenzhen Science \& Technology Fundamental Research Programs (NO: GXWD20201231165807007-20200814115301001 \& JSGG20191129105421211).}}
\address{$^{1}$ADSPLAB, School of ECE, Peking University, Shenzhen, China\\
$^{2}$Peng Cheng Laboratory, Shenzhen, China}
\begin{document}

%
\maketitle
\begin{abstract}
Hand-crafted spatial features, such as inter-channel intensity difference (IID) and inter-channel phase difference (IPD), play a fundamental role in recent deep learning based dual-microphone speech enhancement (DMSE) systems. However, learning the mutual relationship between artificially designed spatial and spectral features is hard in the end-to-end DMSE. In this work, a novel architecture for DMSE using a multi-head cross-attention based convolutional recurrent network (MHCA-CRN) is presented. The proposed MHCA-CRN model includes a channel-wise encoding structure for preserving intra-channel features and a multi-head cross-attention mechanism for fully exploiting cross-channel features. In addition, the proposed approach specifically formulates the decoder with an extra SNR estimator to estimate frame-level SNR under a multi-task learning framework, which is expected to avoid speech distortion led by end-to-end DMSE module. Finally, a spectral gain function is adopted to further suppress the unnatural residual noise. Experiment results demonstrated superior performance of the proposed model against several state-of-the-art models.

\end{abstract}
\section{Introduction}
\label{sec:intro}

Speech communication function of mobile devices has been well-designed and widely used as a convenient tool for contacting others due to its portable characteristics. The quality and intelligibility of the received speech can be severely degraded by background noise if the far-end talker is in an adverse acoustic environment. To attenuate background noise, a two-channel microphone array is typically deployed, where a primary microphone is placed on the bottom of a mobile phone and a secondary microphone on the top.

Conventional dual-channel speech enhancement approaches are based on signal processing, and can be divided into two categories, the blind source separation (BSS) approaches \cite{ref1, ref2}, and the beamforming \cite{ref3, ref4} approaches. Although these conventional approaches are fast and lightweight, their performance and robustness are not reliable in a complex acoustic environment.

Recently, with the success of deep learning based single-channel speech enhancement \cite{ref5,ref5-1}, dual-channel speech enhancement works have been developed on exploring deep learning approaches with conventional speech enhancement methods. One is that the deep neural network (DNN) is used to enhance each microphone signal separately, after which a beamformer is used to linearly integrate the dual-channel signals \cite{ref8, ref9}. Experimental results show that the DNN-beamformer approach gives better results than conventional approaches and shows robustness in terms of various noisy types and SNR ranges. However, the DNN only learns temporal-spectral features of each channel while ignoring the spatial features of the target speech in the DNN-beamformer approach. In order to sufficiently leverage the spatial features of dual-channel data, several approaches have been proposed, which apply the spatial features, such as interaural phase or intensity difference (IPD, IID) \cite{ref6}, as additional inputs for improving objective intelligibility and perceptual quality. Despite performance improvement by learning spectral features together with spatial features, the mutual relationship between spatial and spectral information is difficult to learn by a simple DNN, which may cause the under-utilization of spatial information.

\begin{figure*}[t]
  \centering
  \includegraphics[width=0.7\linewidth]{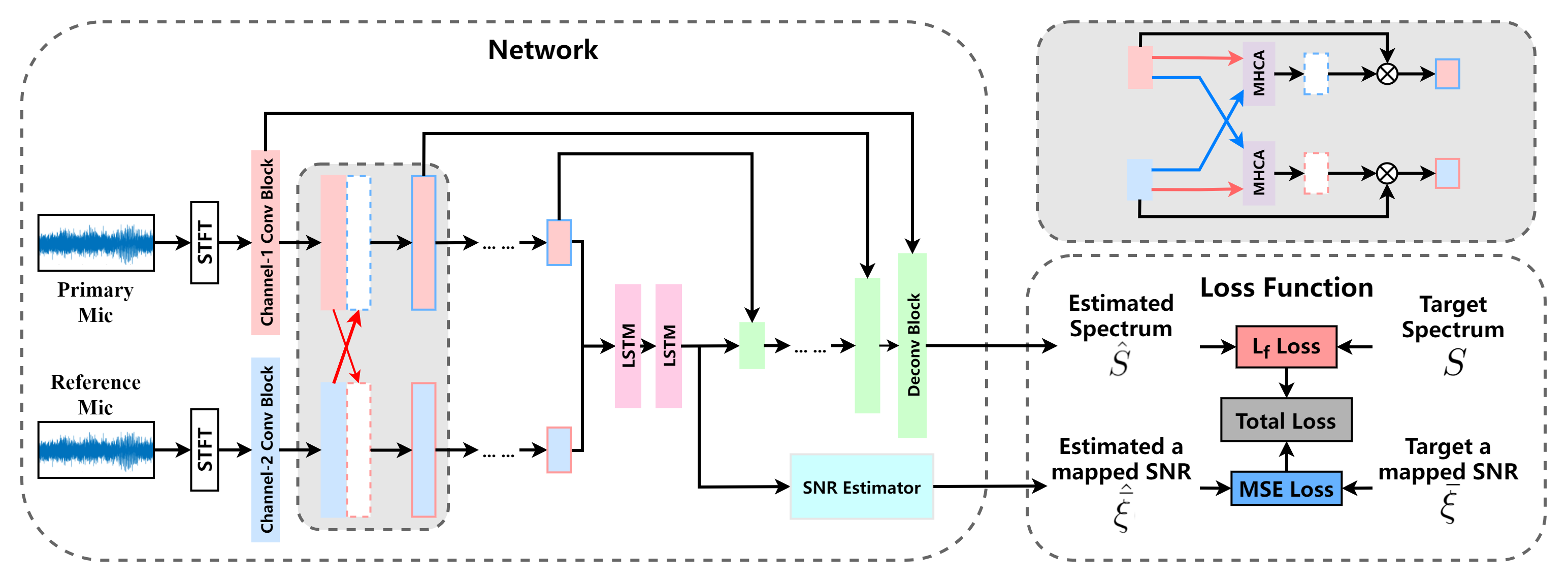}
  \caption{Schematic diagram of MHCA-CRN. The feature maps generated by the encoding layer are interchanged between channels after processing each convolutional block, which is shown in the gray box. Furthermore, MHCA denotes multi-head cross-attention.}
  \label{fig:0}
\end{figure*}

To overcome the aforementioned limitation, this paper is motivated to design a convolutional recurrent network (CRN), which can separately  process each channel for preserving intra-channel features while interchanging information between encoded channels for fully exploiting the spatial information of dual-channel data. For this purpose, this paper proposes a channel-wise encoding structure to process each input channel independently for preserving intra-channel features, and a multi-head cross-attention (MHCA) mechanism to boost network performance by effectively aggregating cross-channel spatial information. In addition, to maintain superior speech quality, the proposed model formulates the decoder as a multi-task learning framework with an auxiliary task of SNR estimation which has proven to be beneficial to the perceived speech quality \cite{ref10}. 

The rest of this paper is organized as follows: the model architecture is presented in Section 2. Section 3 is the dataset and experimental settings. Section 4 demonstrates the results and analysis, and a conclusion is shown in Section 5.

\section{Proposed MHCA-CRN Model}
\label{sec:format}
The proposed MHCA based CRN model (MHCA-CRN) treats the dual-microphone enhancement as a supervised learning task, as shown in Fig.~\ref{fig:0}. First, the proposed model separately encodes the extracted feature from each channel of the noisy signal and interchanges information between encoded channels by using MHCA after each downsampling block. Then the encoded features of both channels are concatenated and fed to LSTM blocks for aggregating temporal contexts. The output of the LSTM blocks is separately fed to the SNR estimator and decoder blocks under a multi-task learning framework. Finally, the output of the SNR estimator block is used to compute a frame-level spectral gain function to remove the residual noise in the estimated spectrum.

\subsection{Encoder-decoder structure}
\label{sec:pagestyle}

The encoder processes each channel independently, in order to preserve the intra-channel feature of dual-channel data and to explicitly utilize the cross-channel relationship. Each encoder contains several stacked 2-D convolutional layers, each of which is followed by batch normalization \cite{ref13-2} and exponential linear unit (ELU) \cite{ref13}. The dilation is applied to the layers along the frequency axis.

The generated feature maps from the encoder of each channel are used as inputs to the MHCA block, which are then interchanged between these two channels. The main objective of the cross-channel attention block is to derive the relationship between two channels.

The decoder is the mirror representation of each encoder except all the convolution layers are replaced with deconvolution layers. Skip connections are introduced to compensate for information loss during the encoding process of the primary microphone.

\subsection{Multi-head cross-attention}
\begin{figure}
  \centering
  \includegraphics[width=0.80\linewidth]{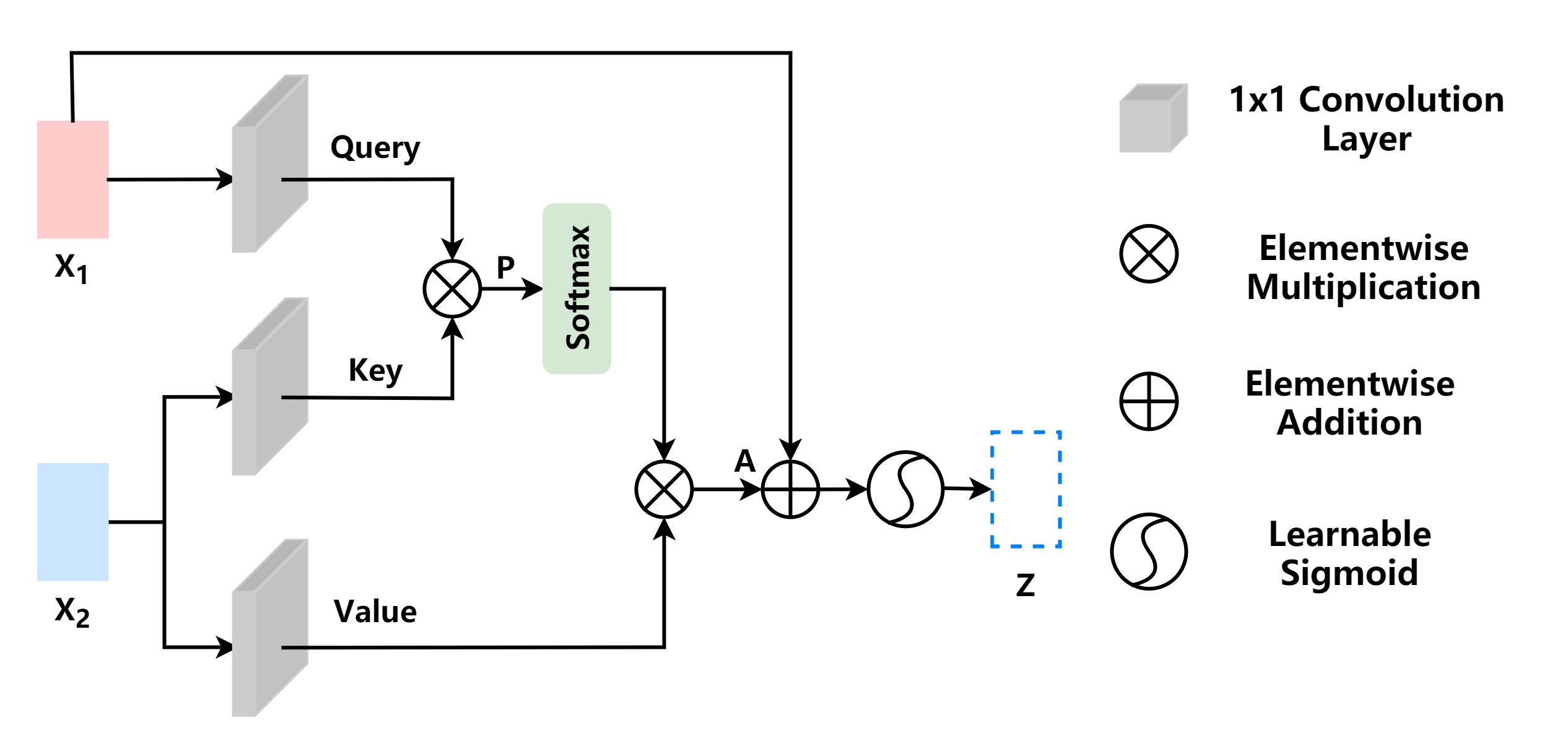}
  \caption{Multi-head cross-attention module.}
  \label{fig:1}
\end{figure}

The MHCA module (shown in Fig~\ref{fig:1}) is designed for synchronizing time delay between two channels, which holds the spatial information of the target speaker. The MHCA takes transformed feature maps corresponding to the encoder of primary channel, $\textbf{X}_1$, processed a $1\times1$-convolution block to form the query, and takes transformed feature maps of the encoder of reference channel, $\textbf{X}_2$, to form the key-value pair by using two $1\times1$-convolution blocks. The proposed MHCA first computes the query, and key-value pair for obtaining the attention component \textbf{A}, given by
\begin{equation}
    \textbf{A}=\text{softmax}(\textbf{Q} \textbf{K}^{\top}) \textbf{V}
\end{equation}
where \textbf{Q}, \textbf{K}, and \textbf{V} denote the query, key, and value, respectively. Intuitively, the multiplication operation between \textbf{Q} and \textbf{K} emphasizes the regions which are slowly varying in time and have high power. What is more, the output of MHCA block \textbf{Z} is computed by:
\begin{equation}
  \textbf{Z} = \text{sigmoid}(\textbf{X}_1 + \textbf{A})
\end{equation}
Consequently, the \textbf{Z} is weight value that are re-scaled between 0 and 1 through a sigmoid activation function.

\subsection{SNR estimator}
Previous researches have proved that directly training DNN models may inevitably cause a certain amount of speech distortion \cite{ref13-3}. To this end, we propose to utilize SNR estimator to estimate frame-level SNR under a multi-task learning framework for maintaining speech quality while reducing noise.

The input of the SNR estimator is the feature maps obtained by two LSTM layers, then it is fed to a convolution layer with sigmoid activation to estimate the frame-level SNR. The training target for the SNR estimator is the mapped SNR \cite{ref14}, which is a mapped version of instantaneous SNR. The definition of instantaneous SNR is as follows:
\begin{equation}
    \xi_{dB}(t,f) = 20\log_{10}(|S(t,f)|/|N_1(t,f)|)
\end{equation}
where $t$ and $f$ denote the time and frame index, $\xi_{dB}(t,f)$ is scaled in [0, 1] and can be viewed as \textit{a priori} SNR, $S(t,f)$ and $N_1(t,f)$ are respectively the clean and noise spectrum of the primary channel. In addition, it is assumed that $\xi_{dB}(t,f)$ is distributed normally with mean, $\mu_f$, and variance, $\sigma^2_f$: $\xi_{dB}(t,f)\sim \mathcal{N}(\mu_f, \sigma^2_f)$. The mapped SNR is given by
\begin{equation}
    \bar{\xi}(t,f)=1+\Big[\frac{1}{2}+\text{erf}\Big(\frac{\xi_{dB}(t,f)-\mu_f}{\sigma_f\sqrt{2}}\Big)\Big]
\end{equation}
where ``erf'' is the error function.

During inference (shown in Fig~\ref{fig:2}), the SNR estimate, $\hat{\xi}(t,f)$ is computed by
\begin{equation}
    \hat{\xi}(t,f) = 10^{((\sigma_f\sqrt{2}\text{erf}^{-1}(2\hat{\bar{\xi}}(t,f)-1)+\mu_f)/10)}
\end{equation}
where $\hat{\xi}(t,f)$ is output from SNR estimator.
\subsection{Loss function}
Since proposed MHCA-CRN model formulates the decoder with an extra SNR estimator under a multi-task learning framework, the proposed MHCA-CRN model is trained by a combination of two losses. First, the MSE loss to guide the learning of SNR estimator,
\begin{equation}
    L_{SNR}=\text{MSE}(\xi(t,f), \hat{\xi}(t, f))
\end{equation}

Second, the loss \cite{ref15-1} for target speech spectrum reconstruction is given by:
\begin{equation}
\begin{split}
    L_f=\frac{1}{T\times F}\sum_{t=1}^T\sum_{f=1}^{F}|(|S(t,f)|)
    -(|\hat{S}(t,f)|)|
\end{split}
\end{equation}

Finally, the total loss is
\begin{equation}
    L=  L_f + \alpha L_{SNR}
\end{equation}
Since both loss values are not on the same scale, we empirically set $\alpha$ to 10.

\subsection{Target speech reconstruction}
\label{sec:majhead}

For further suppressing the residual noise, the proposed model adopts the spectral gain function , $G^{SNR}(t,f)$, for dual-microphone speech enhancement, which is represented as
\begin{equation}
    G^{SNR}(t,f) = \frac{\hat{\xi}(t,f)}{\hat{\xi}(t,f)+1}
\end{equation}
where $\hat{\xi}(t,f)$ is the estimated SNR obtained by SNR estimator. 

\begin{figure}
  \centering
  \includegraphics[width=0.80\linewidth]{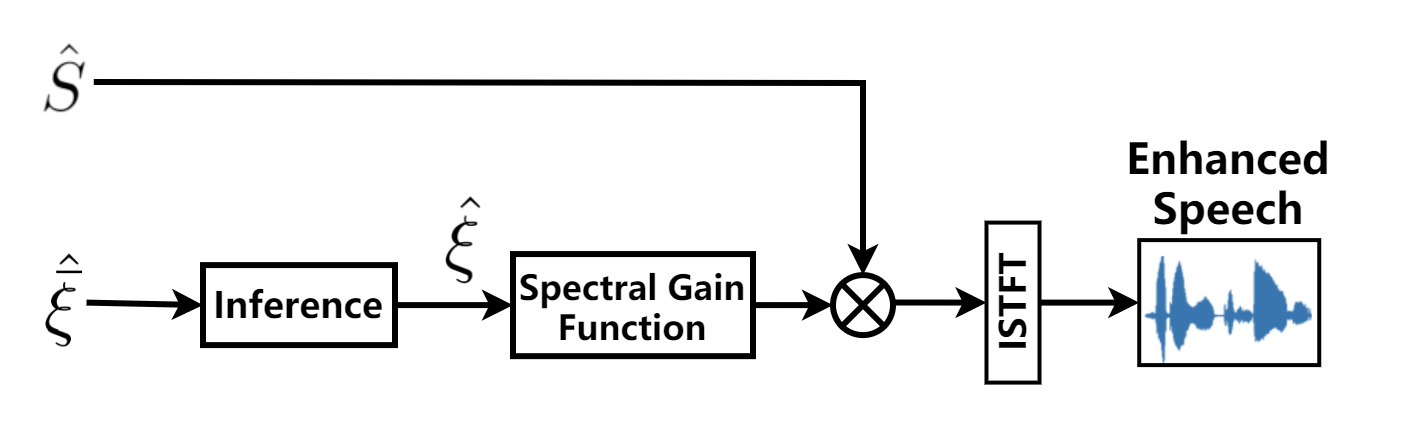}
  \caption{Diagram of target speech reconstruction.}
  \label{fig:2}
\end{figure}

The computed final gain $G^{SNR}(t,f)$ is then multiplied to the estimated spectrum $\hat{S}(t,f)$ to suppress the residual noise, which is then combined with the noisy phase to resynthesize the time-domain waveform of the enhanced speech, as shown in Fig~\ref{fig:2}. 

\renewcommand{\arraystretch}{0.85}
\begin{table*}[]
\centering
\caption{PESQ and STOI comparison for the different models. A higher score means better performance where the \textbf{BOLD} text indicates the best performance for each metric.}
\begin{tabular}{|l|c|cc|cc|cc|cc|}
\hline
Test SNR        & Channel & \multicolumn{2}{c|}{-5 dB}     & \multicolumn{2}{c|}{0 dB}      & \multicolumn{2}{c|}{5 dB}      & \multicolumn{2}{c|}{10 dB}     \\ \hline
Meric           & -            & STOI(\%)       & PESQ          & STOI(\%)       & PESQ          & STOI(\%)       & PESQ          & STOI(\%)       & PESQ          \\ \hline
Unprocessed     & Dual            & 57.06          & 1.37          & 69.33          & 1.88          & 80.59          & 2.12          & 87.63          & 2.49          \\ \hline
DeepXi \cite{ref14}         & Single            & 77.48          & 1.88          & 90.27          & 2.36          & 92.17          & 2.66          & 95.74          & 3.11          \\ \hline
CB-NR \cite{ref19}           & Dual            & 54.42          & 1.46          & 68.31          & 2.03          & 77.57          & 2.38          & 88.10          & 2.74          \\
CRN-PSM \cite{ref6}        & Dual            & 78.20          & 1.76          & 87.30          & 2.17          & 92.76          & 2.59          & 95.76          & 2.99          \\
DC-CRN \cite{ref7}         & Dual            & 86.54          & 2.48          & 92.64          & 2.94          & 95.88          & 3.20          & 97.47          & 3.43          \\ \hline
MHCA-CRN  & Dual            & \textbf{86.58} & \textbf{2.51} & \textbf{92.83} & \textbf{3.03} & \textbf{95.96} & \textbf{3.24} & \textbf{97.52} & \textbf{3.46}          \\
\quad -without spectral mapping  & Dual            & 84.98          & 2.17          & 89.80          & 2.57          & 93.82          & 2.92          & 95.47          & 3.19          \\
\quad -without SNR estimator & Dual      & 85.76     & 2.39          & 92.34         & 2.88      & 95.47         & 3.11      & 97.02        & 3.38          \\ 
\quad -without MHCA blocks & Dual      & 82.93     & 2.08          & 90.06         & 2.64      & 94.16         & 2.98      & 95.84        & 3.16          \\\hline

\end{tabular}
\end{table*}

\section{Experimental setup}
\label{sec:print}
\subsection{Data preparation}
29 hours and 1 hour of speech are selected from Librispeech corpus as training and validation sets, respectively. The noises are from the DEMAND dataset. In addition, this paper simulate room impulse response (RIR) by the IMAGE method \cite{ref18}. Specifically, two microphones with 2cm interval are placed at the center of a $5m$(length) $\times$ $5m$(width) $\times$ $3m$(height) room, and noise source which are placed at 1.5m away from the center of the two microphones and ranged from $0^{\circ}$ to $360^{\circ}$ spaced by $10^{\circ}$. 

For each mixture, a speech and a slice of noise are randomly chosen and are placed at two different positions, and the speech and noise are mixed under the randomly SNR levels ranging from -5dB to 10dB. In addition, the frame length is 32 ms and the hop size is 16 ms. The Hanning window is used as the analysis window. The sampling rate is 16 kHz. A 512-point discrete Fourier transform is used to extract complex short-time Fourier transform (STFT) spectrograms.

\subsection{Baselines and training details}
\label{sec:page}
The proposed MHCA-CRN model has been compared with four other baselines: (1) \textbf{CB-NR} \cite{ref19}: A coherence-based dual-channel noise reduction algorithm; (2) \textbf{DeepXi} \cite{ref14}: A minimum mean-square error (MMSE) approach for single-channel speech enhancement by using deep learning; (3) \textbf{CRN-PSM} \cite{ref6}: a CRN approach to predict phase sensitive mask (PSM) for dual-microphone speech enhancement; (4) \textbf{DC-CRN} \cite{ref7}: a densely-connected CRN approach for mobile communication based on dual-channel complex spectral mapping.

To better validate the proposed structure and strategies, we add three ablation experiments. Firstly, we remove the SNR estimator and keep deconvolution layers to predict the clean speech spectrum directly. Secondly, we keep the SNR estimator only and remove the deconvolution layers for comparing the performance between single channel deep learning based MMSE approach, \textit{i.e.} DeepXi, and dual-channel deep learning based MMSE approach. Finally, we remove the MHCA blocks to evaluate the effectiveness of MHCA. 

For the training step, all models are trained with Adam optimizer for stochastic gradient descent (SGD) based optimization. The learning rate is set to 0.001. All training samples are zero-padded to have the same number of time steps as the longest sample.

\begin{figure}[t]
	\centering
	\includegraphics[width=0.15\textwidth]{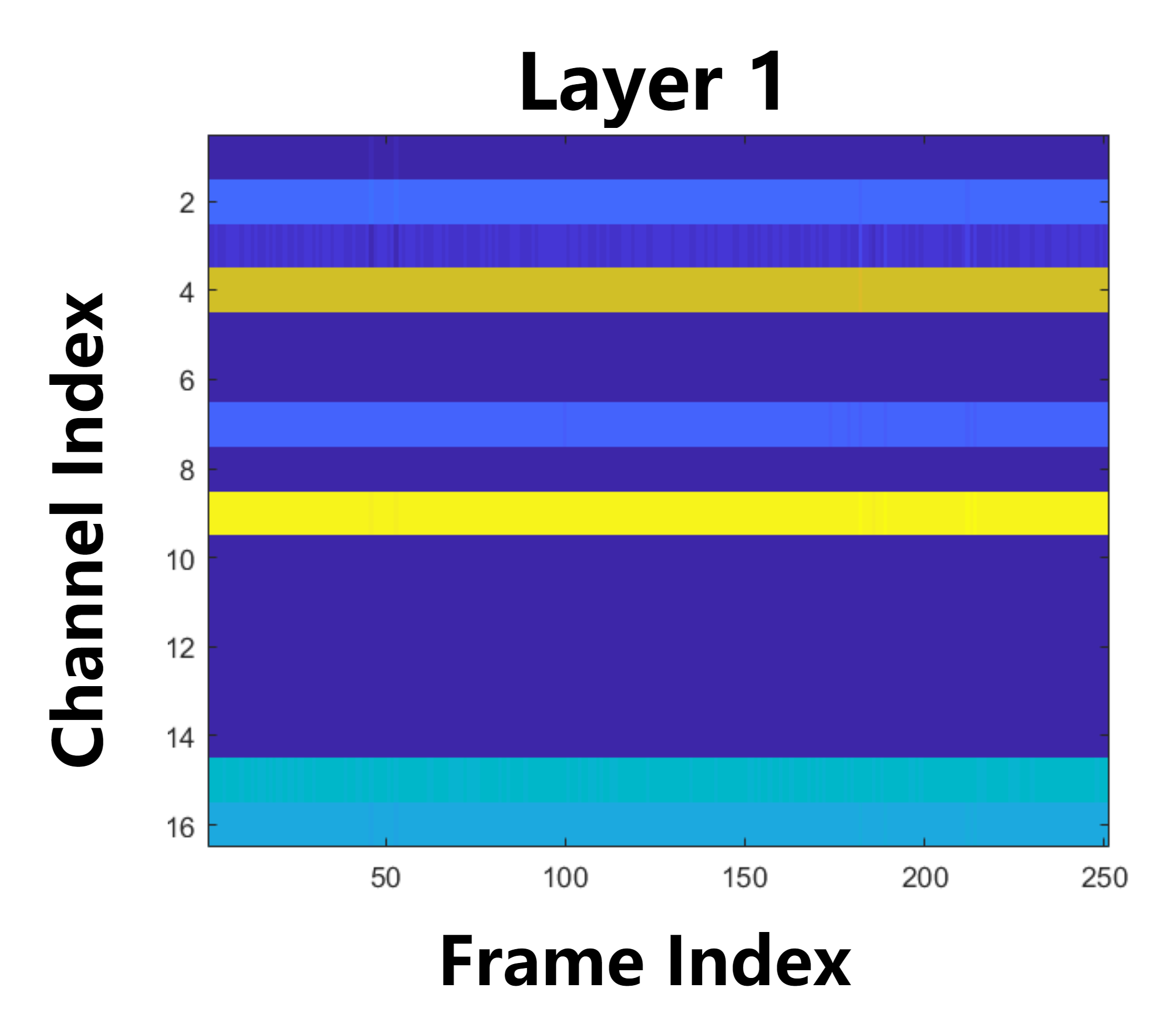} 
	\includegraphics[width=0.15\textwidth]{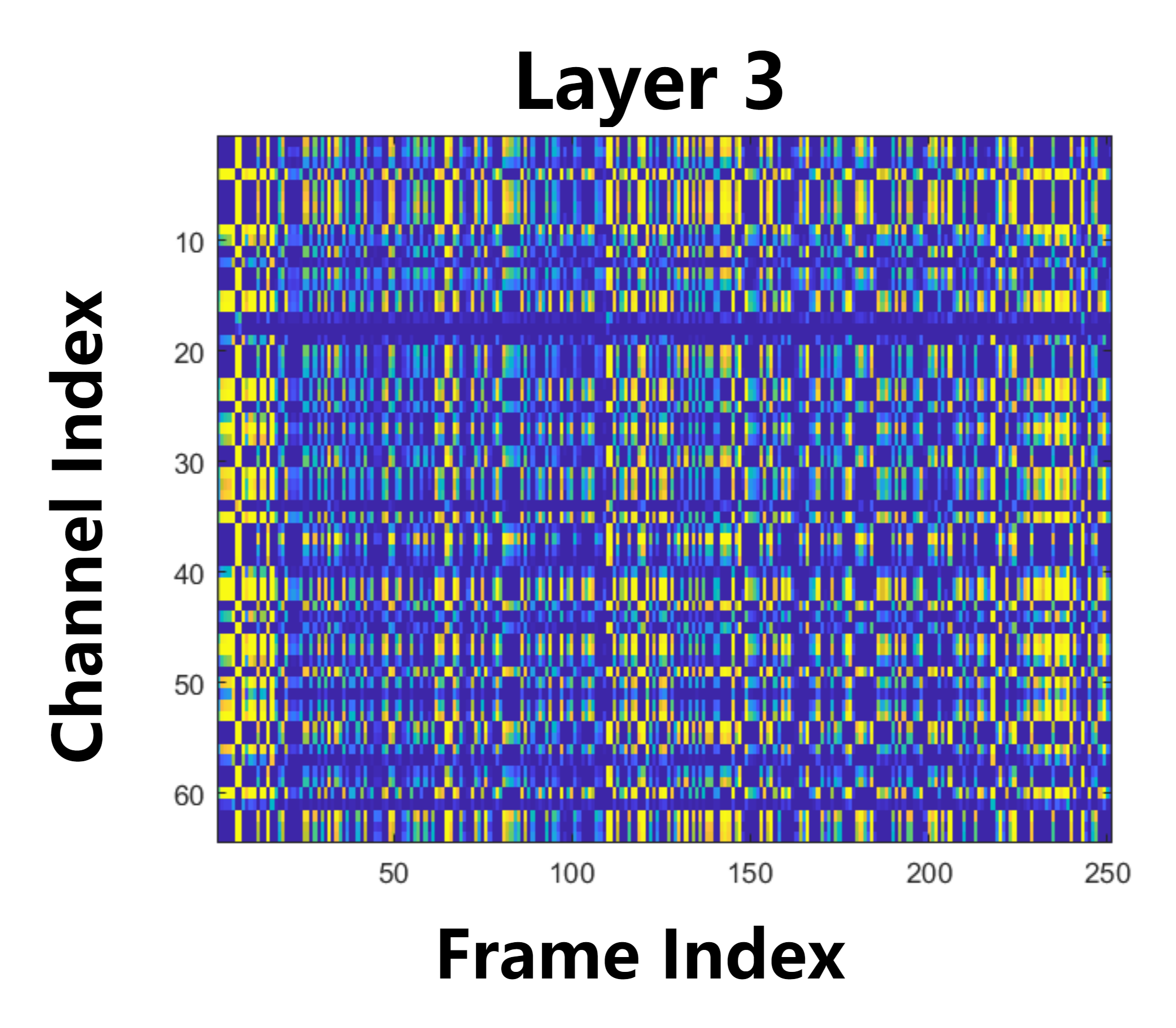}
	\includegraphics[width=0.15\textwidth]{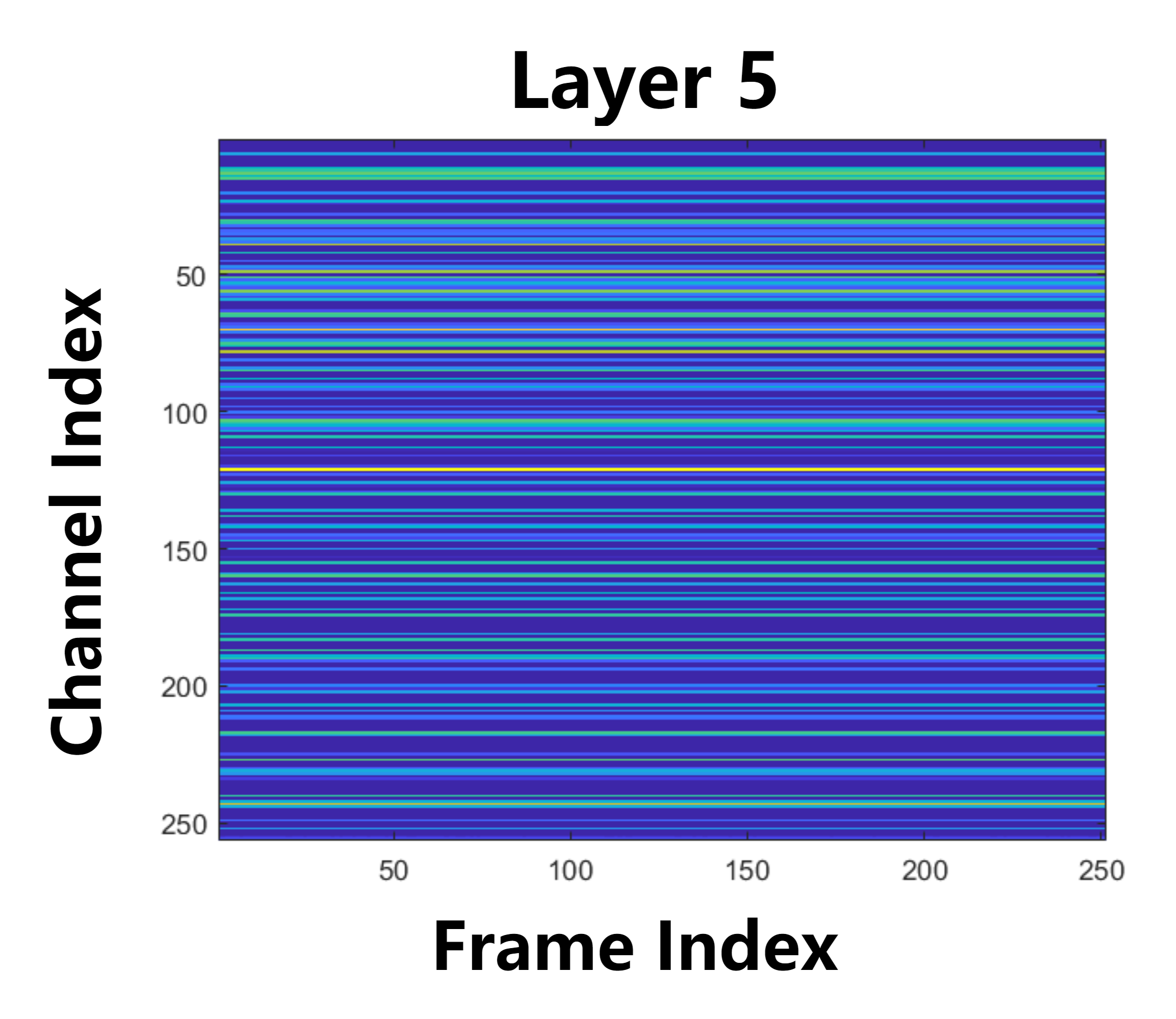}
	\caption{From left to right: Visualization of MHCA mask at layer 1, layer 3, and layer 5. }
	\label{fig:3}
\end{figure}

\section{Results and analysis}
The speech enhancement systems are evaluated using Perceptual Evaluation Speech Quality (PESQ) and Short Term Objective Intelligibility (STOI). Experimental results are summarized in Table 1.

According to Table 1, we have the following observations: Firstly, the deep learning based methods significantly improve both STOI and PESQ metrics, and outperform the conventional approach, \textit{i.e.}, CB-NR. Secondly, the MHCA-CRN without spectral mapping performs better than DeepXi in terms of PESQ and STOI, which indicates that the dual-channel based model show more robustness than single-channel based model. Finally, apart from that, the proposed MHCA-CRN achieves better results than MHCA-CRN without SNR estimator, the proposed MHCA-CRN consistently outperforms the state-of-the-art model, \textit{i.e.}, DC-CRN, in both metrics. This indicate the effectiveness of the SNR estimator.

Table 1 furthermore shows that the MHCA has significantly improved network performance. The visualization of MHCA masks at different convolution layers is as shown in Fig~\ref{fig:3}. In early layers, the masks pay more attention to certain feature map channels, for example, the $4^{th}$ and $9^{th}$ channels of the feature map are highlighted by the mask at the first layer. What is more, when the layer is going deeper, the shape of the MHCA mask is changing in order to synchronize the time delay between two channels. Take, for example, mask at $3^{rd}$ layer highlights several channels but at different time frames, while mask at $7^{th}$ highlights certain channels at all time frames. This is demonstrated that the spatial cues between dual-channel can be implicitly exploited by MHCA. 

\section{Conclusion}
\label{sec:refs}
In this paper, we propose an MHCA-CRN for dual-microphone speech enhancement, aiming to straightforwardly and efficiently exploit spatial information. The model adopts a channel-wise encoding structure to process each input channel independently for preserving intra-channel features and uses the MHCA mechanism to aggregate cross-channel spatial information. Furthermore, an SNR estimator is adopted along with the decoder to estimate frame-level SNR under a multi-task learning framework for further improving the speech quality. Finally, a spectral gain function is adopted to remove unnatural residual noise. Experimental results show that our proposed method can suppress the noise meanwhile maintaining better intelligibility.

\vfill\pagebreak
\bibliographystyle{IEEEbib}
\bibliography{refs}

\end{document}